# Mapping the Ages of Stars with Chemistry

*ESO White Paper*

**Carlos Viscasillas Vázquez**

*INAF – Osservatorio Astrofisico di Arcetri, Largo E. Fermi 5, 50125, Firenze, Italy*

*Vilnius University, Faculty of Physics, ITPA, Sauletekio av. 3, 10257 Vilnius, Lithuania*

**Giada Casali**

*INAF – Osservatorio di Astrofisica e Scienza dello Spazio, via P. Gobetti 93/3, 40129 Bologna, Italy*

*Research School of Astronomy and Astrophysics, The Australian National University, Canberra, ACT 2611, Australia*

**Laura Magrini**

*INAF – Osservatorio Astrofisico di Arcetri, Largo E. Fermi 5, 50125, Firenze, Italy*

**Gabriele Cescutti**

*Dipartimento di Fisica, Sezione di Astronomia, Università di Trieste, Via G. B. Tiepolo 11, 34143 Trieste, Italy*

**Sergio Cristallo**

*INAF – Osservatorio Astronomico d'Abruzzo, via Mentore Maggini snc, I-64100, Teramo, Italy*

*INFN – Sezione di Perugia, via Alessandro Pascoli snc, I-06123, Perugia, Italy*

**Camilla Danielski**

*Departament d'Astronomia i Astrofísica, Universitat de València, C. Dr. Moliner 50, 46100 Burjassot, Spain*

*INAF – Osservatorio Astrofisico di Arcetri, Largo E. Fermi 5, 50125, Firenze, Italy*

**Riano Giribaldi**

*INAF – Osservatorio Astrofisico di Arcetri, Largo E. Fermi 5, 50125, Firenze, Italy*

**Georges Kordopatis**

*Université Côte d'Azur, Observatoire de la Côte d'Azur, CNRS, Laboratoire Lagrange, Nice, France*

**Ivan Minchev**

*Leibniz-Institut für Astrophysik Potsdam (AIP), An der Sternwarte 16, D-14482, Potsdam, Germany*

**Marta Molero**

*Institut für Kernphysik, Technische Universität Darmstadt, Schlossgartenstr. 2, Darmstadt 64289, Germany*

*INAF, Osservatorio Astronomico di Trieste, Via Tiepolo 11, 34131 Trieste, Italy*

**Andrés Moya**

*Departament d'Astronomia i Astrofísica, Universitat de València, C. Dr. Moliner 50, 46100 Burjassot, Spain*

**Marco Palla**

*INAF – Osservatorio Astrofisico di Arcetri, Largo E. Fermi 5, 50125, Firenze, Italy*

**Gražina Tautvaišienė**

*Vilnius University, Faculty of Physics, ITPA, Sauletekio av. 3, 10257 Vilnius, Lithuania*

**Diego Vescovi**

*INAF – Osservatorio Astronomico d'Abruzzo, via Mentore Maggini snc, I-64100, Teramo, Italy*

*INFN – Sezione di Perugia, via Alessandro Pascoli snc, I-06123, Perugia, Italy*

## 1. Introduction

Stellar ages are fundamental to nearly all branches of astrophysics, from Galactic archaeology to planet formation and cosmology, but they remain among the most difficult stellar parameters to measure. Classical techniques such as isochrone fitting lose precision for long-lived evolutionary phases, including the main sequence and the red giant branch, while asteroseismic constraints, although extremely powerful, remain limited to a small subset of the Milky Way's stellar populations. In this context, chemical clocks, namely abundance ratios whose evolution is sensitive to stellar age, have emerged as a promising approach. Notable examples include [*s*-process/α-element] ratios such as [Y/Mg] and [Ce/Ti], the [C/N] ratio in low and intermediate-mass red giant branch stars affected by first dredge-up and extra mixing, and the depletion of lithium in young late-type dwarfs. Each of these indicators traces a distinct physical mechanism: Galactic chemical enrichment, the internal structural evolution of stars, or the depletion and mixing of fragile elements in stellar envelopes, and together they offer the possibility of inferring ages across a wide variety of evolutionary phases and Galactic environments. Great expectations belong to the high-resolution spectroscopy of carbon isotope ratios that are very sensitive to stellar ages in evolved low-mass giants. Looking ahead to the 2040s, the combination of high-precision spectroscopy, next-generation asteroseismology, and deep, all-sky astrometry will provide, for the first time, the capability to unify these diverse clocks into a coherent and multi-dimensional mapping of stellar ages throughout the Milky Way.

## 2. Key Science Questions for the 2040s and Beyond

### 2.1. Can we build a unified, environment-dependent calibration of chemical clocks across the Galaxy?

The widely used [Y/Mg]-age relation since the pioneering works of da Silva et al. (2012) and Nissen (2015) shows significant variation across the different components of the Milky Way. It is essentially flat in the thick disk (Tautvaišienė et al. 2021), reflecting the rapid early formation of this population dominated by Type II supernova enrichment, while in the thin disk it displays a clear negative slope driven by the delayed contribution of yttrium from asymptotic giant branch (AGB) stars. Spatial trends across the disk further complicate the picture: the relation flattens and shifts to lower [*s*-/α] in the inner Galaxy, and becomes steeper and offset to higher [*s*-/α] toward the outer disk (Viscasillas Vázquez et al. 2022). This trend is even more pronounced when expressed in terms of stellar birth radius (Ratcliffe et al. 2024), reflecting differences in local star formation histories, the inside-out formation of the Galactic disk, and chemical enrichment timescales. Although recent chemical-evolution models reproduce some of these features (Magrini et al. 2021), they still fail to capture the observed patterns in the inner Galaxy towards young ages (Molero et al. 2025), indicating that key physical ingredients, such as metallicity-dependent AGB yields, radial migration, or multi-phase gas accretion, are still incomplete and/or inaccurately parameterized. This leads to one of the central questions for the 2040s: whether it is possible to construct a truly universal, Galaxy-wide chemical clock, or whether stellar ages must ultimately be inferred through multi-parameter, environment-dependent calibrations. Addressing this challenge requires increases in sample size, abundance precision, and spatial coverage, well beyond the capabilities of current and near-term spectroscopic facilities.

### 2.2. How can internal-evolution clocks like C/N be anchored to absolute ages?

The [C/N] ratio in red giant stars is modified during the first dredge-up in a way that depends on stellar mass, and because mass is a strong proxy for age in evolved stars. Theoretically, the [C/N] versus age relations were first modelled by Salaris et al. (2015) for first dredge-up giants and later by Lagarde et al. (2017) for red-clump stars. Yet predictions of the dredge-up from stellar models alone remain insufficiently accurate to deliver absolute ages, making external calibration essential. The most reliable benchmarks come from open clusters with well-determined ages (Tautvaišienė et al. 2025) and from red giants with asteroseismic masses and ages (Roberts et al 2025) obtained from missions such as Kepler, K2, TESS, and, in the 2030s and 2040s, PLATO and the proposed HAYDN mission. The tight correlation between [C/N] and seismic mass established in the APOKASC sample has demonstrated the potential of this clock, and new measurements from SDSS-IV and SDSS-V have now extended empirical calibrations into the metal-poor regime, where "metal-poor" refers specifically to the low-metallicity tail of the Galactic disk, and not to genuinely metal-poor or halo-like stellar populations. A central question for the 2040s is whether these calibrations can be expanded to encompass the full extent of the Galactic disk, bulge, and halo while accounting for the effects of extra mixing, metallicity, and mass loss that may vary across different stellar environments. Meeting this goal will require high-resolution spectroscopy (R > 40,000) over wide fields and reaching faint magnitudes, enabling precise [C/N] measurements for vast numbers of stars throughout the Milky Way.

### 2.3. How do multiple clocks combine to produce a robust multi-dimensional age estimator?

Different chemical clocks trace distinct layers of stellar physics: ratios such as [Y/Mg], [Y/Al], or [Ba/Mg], and more recently, [Ce/Mg] and [Zr/Ti] (Casali et al. 2025), reflect the competition between AGB enrichment and Type II supernova yields and encode the timescales of Galactic chemical evolution; the [C/N] ratio in red giants probes internal mixing processes that depend on stellar mass and evolutionary state; and lithium depletion for relatively young stars traces envelope mixing and the gradual destruction of fragile elements. In Moya et al. (2022), a first machine-learning (ML) model combining different chemical clocks in a hierarchical Bayesian framework was presented, showing what can be achieved compared to using a single chemical clock. In Tamames-Rodero et al. (2025), this model was improved using a more flexible Hierarchical Bayesian Neural Network (HBNN) for this combination. This will allow the inclusion of other dependences in the model, such as the Galactocentric present-day radius and birth radius, as described in Sec. 2.1. For all these ML models, a large and precise data set is essential for assessing a precise age estimation. With the combination of high-resolution, high-multiplex spectroscopic capabilities and complementary asteroseismic and open-cluster benchmarks, the 2040s may enable the simultaneous measurement of multiple classes of chemical age tracers for tens of millions of stars, allowing their consistent calibration and their combination through advanced ML techniques to derive multi-dimensional chemical ages that surpass the limitations of any single indicator. Realizing this vision, however, will require continuous, homogeneous, high-resolution spectroscopy over enormous areas of sky, a capability that lies beyond what the ELT or any existing or planned multiplexed facility of the 2030s can provide.

### 2.4. What do chemical clocks reveal about the structure and formation history of the Milky Way?

In the 2040s, the application of chemical clocks will make it possible to probe the number and timescales of disk-infall episodes (e.g. Palla et al. 2024), to reconstruct the age structure of the bar and bulge, to quantify the efficiency of radial migration as a function of Galactocentric radius and stellar age (e.g., Minchev et al. 2014, Viscasillas Vázquez et al. 2023, Kordopatis et al. 2025), to assess the delayed formation history of the outer disk, and to trace the Galactic halo accretion history encoded in the joint distribution of ages, chemical abundances, and stellar kinematics. In planetary science, determining stellar ages provides the essential temporal framework for understanding how long planetary systems have had to evolve. This, in turn, helps constrain the pathways that produced the architectures we observe today. Combined with stellar chemical compositions and kinematics, such ages will deepen our understanding of the Galaxy's planetary population, a timely need given the large number of new planets expected from Gaia, the Nancy Grace Roman Space Telescope, and PLATO. To unlock this potential, it will be essential to obtain precise abundance measurements for age-sensitive tracers across the entire Galactic system, including the entire disk, the bulge and bar, the stellar halo, and the diffuse, low-surface-brightness structures at large radii. Meeting these requirements demands observational capabilities far beyond those expected to be available by the end of the 2030s.

### 3. Why a New Facility Is Essential Beyond the 2030s

High-resolution spectroscopy on a scale of tens to hundreds of millions of stars will be essential for exploiting the full diagnostic power of chemical clocks. Achieving this requires resolutions above R~40,000 to measure several *s*-process elements, such as Y, Zr, Sr, Ba, Ce, La, many α-elements, as, e.g., Ti, Mg, and light elements C, N, O, with the precision needed for age determination (Kordopatis et al. 2023b), along with broad wavelength coverage from roughly 350 to 970 nm to access both the blue $s$-process lines and the red CN bands. Abundance precision at the level of a few hundredths of a dex and sensitivity down to $G \sim 19$ mag are also necessary. As an illustrative example, using the relations of [Casali et al. (2020)](#) for Solar twins in the solar neighborhood, measurement uncertainties of 0.01 dex in A(Y), A(Mg), and A(Fe) translate into typical age errors of about 0.5 Gyr. These uncertainties increase to roughly 0.9 Gyr when assuming larger abundance errors of 0.02 dex, and 1.3 Gyr for errors of 0.03 dex. Clearly, we need to consider also the systematic errors in the total budget, which might increase the final uncertainty on the age. However, the large statistics that can be achieved with upcoming high-resolution spectroscopic surveys will help mitigate part of these limitations by enabling a robust characterization of the intrinsic scatter and a better control of systematic effects. Existing multiplexed facilities such as WEAVE, 4MOST, or MOONS do not reach the required combination of resolution, throughput, and collecting area, while the ELT will deliver exquisite precision for individual targets but lacks the multiplexing needed for Galactic-scale chemo-chronology. To anchor chemical clocks to physical ages, an all-sky, high-cadence asteroseismic space mission will also be needed in the 2040s. Such a facility must deliver seismic ages for large samples of stars across the Galaxy, improve upon the precision and target density expected from PLATO, and be capable of observing crowded cluster fields, as envisioned for missions like HAYDN. These seismic constraints are indispensable for calibrating both empirical and physically motivated age relations, providing the absolute age scale against which chemical tracers must be tied. Meeting the scientific goals of the 2040s will furthermore require exascale data pipelines and ML frameworks capable of integrating diverse spectroscopic, seismic, and astrometric information. This includes NLTE and 3D spectral modeling applied at scale, AI-driven forward-modeling architectures, and the multi-dimensional combination of abundance ratios such as Y/Mg, C/N, Li, and α/Fe together with seismic ages. Long-term, standardized cross-calibration across instruments, surveys, and decades of observing will be essential to ensure coherence and accuracy in the resulting Galactic age map.

## 4. Future Technology and Data Requirements

The science outlined in this white paper demands the development of a new wide-field, high-resolution spectroscopic facility, an instrument on a >10 m-class telescope with thousands of fibers at resolving power above $R \approx 40{,}000$, capable of delivering the wavelength coverage and sensitivity required to measure weak spectral lines of key age-sensitive elements, particularly in the UV and blue spectral regions ([Kordopatis et al. 2023b](#)). Its impact will rely on strong synergy with the next generation of space-based asteroseismic missions, such as an enhanced PLATO or the proposed HAYDN mission, which together will provide the seismic benchmarks needed to anchor chemical age scales across the Galaxy. Extracting the full scientific return from such facilities will also require exascale data-processing infrastructures designed to merge spectroscopic, seismic, and astrometric information into a coherent framework, supported by long-term cross-calibration strategies that link cluster ages, seismic ages, and chemically inferred ages on a unified scale. ML approaches will play a central role in this process, enabling the combination of dozens of elemental abundance ratios into precise and internally consistent age estimates. In particular, the Bayesian Stacking of a set of different ML models such as HBNN, Bayesian Random Forest, or Gaussian Process is found to be the most efficient option for stellar dating using different chemical clocks, but this model is very data and computing-demanding. Taken together, these advances represent natural and necessary technological developments for the astrophysical landscape of the 2040s. A facility such as the planned 12-m Wide-Field Spectroscopic Telescope (WST, [Mainieri et al. 2024](#)), with its unprecedented combination of high resolution, wide field, and massive multiplexing, is uniquely suited to deliver the homogeneous, precision spectroscopy required to calibrate and apply chemical clocks across the entire Milky Way. Its scale and survey speed make it the only planned observatory capable of providing the millions of high-quality spectra needed to realise a fully multi-dimensional age map in the 2040s.